\documentclass[prx,twocolumn,floatfix,superscriptaddress,longbibliography]{revtex4-1}

\usepackage{amssymb,amsmath,amstext}
\usepackage{graphicx}
\usepackage{epstopdf}
\usepackage{color}
\usepackage{bm}
\usepackage{appendix}
\usepackage[T1]{fontenc}
\usepackage{bbold}
\usepackage{bbm}
\usepackage{latexsym}
\usepackage[colorlinks=true,citecolor=blue,linkcolor=magenta]{hyperref}
\usepackage{float}
\usepackage{braket}

\graphicspath{{Figures/}}

\newcommand{\edit}[1]{{\color{blue}{{#1}}}}

\begin{document}

\title{Robust design under uncertainty in quantum error mitigation}

\author{Maksym Prodius} 
\thanks{The first two authors contributed equally to this work.}
\affiliation{Institute of Theoretical Physics, Jagiellonian University, Krakow, Poland.}
\affiliation{Szko\l{}a Doktorska Nauk \'Scis\l{}ych i Przyrodniczych, Uniwersytet Jagiello\'nski, ulica Stanis\l{}awa \L{}ojasiewicza 11, PL-30-348 Krak\'ow, Poland}

\author{Piotr Czarnik} 
\thanks{The first two authors contributed equally to this work.}
\affiliation{Institute of Theoretical Physics, Jagiellonian University, Krakow, Poland.}
\affiliation{Mark Kac Center for Complex Systems Research, Jagiellonian University, Krak\'ow, Poland}

\author{Michael McKerns} 
\affiliation{Information Sciences, Los Alamos National Laboratory, Los Alamos, NM, USA.}
\affiliation{The Uncertainty Quantification Foundation, Wilmington, DE 19801, USA.}

\author{Andrew T. Sornborger} 
\affiliation{Information Sciences, Los Alamos National Laboratory, Los Alamos, NM, USA.}
\affiliation{Quantum Science Center, Oak Ridge, TN 37931, USA.}

\author{Lukasz Cincio}
\affiliation{Theoretical Division, Los Alamos National Laboratory, Los Alamos, NM, USA.}
\affiliation{Quantum Science Center, Oak Ridge, TN 37931, USA.}

\begin{abstract}
Error mitigation techniques are crucial to achieving near-term quantum advantage. Classical post-processing of quantum computation outcomes is a popular approach for error mitigation, which includes methods such as Zero Noise Extrapolation, Virtual Distillation, and learning-based error mitigation. However, these techniques have limitations due to the propagation of uncertainty resulting from the finite shot number of a quantum measurement.
In this work, we introduce general and unbiased methods for quantifying the uncertainty and error of error-mitigated observables based on the strategic sampling of error mitigation outcomes. We then extend our approach to demonstrate the optimization of performance and robustness of error mitigation under uncertainty.
To illustrate our methods, we apply them to Zero Noise Extrapolation and Clifford Date Regression in the ground state of the XY model simulated using depolarizing and IBM Toronto noise models, respectively.
In particular, we optimize the choice of  noise levels and the allocation of shots for Zero Noise Extrapolation and the distribution of the training circuits for Clifford Data Regression.
While our methods are readily applicable to any post-processing-based error mitigation approach, in practice they must not be prohibitively expensive -- even
though they perform optimizations of the error mitigation hyperparameters requiring sampling of a statistical distribution of error mitigation outcomes.
By leveraging surrogate-based optimization \cite{o2025efficient}, we show
that our methods can efficiently perform optimal design for a Zero Noise Extrapolation
implementation. We then further demonstrate the transferability of learned Zero Noise Extrapolation hyperparameters to other similar circuits.

\end{abstract}
\maketitle

\section{Introduction}

Quantum computers promise to outperform the best classical computers. Such quantum advantage has already been claimed for some tasks~\cite{google2019supremacy,morvan2023phase,kim2023evidence}. Nevertheless, the potential of current gate-based quantum computers is severely limited due to decoherence and imperfect implementations of quantum gates, so-called hardware noise~\cite{cerezo2020variationalreview,endo2021hybrid}. It is commonly expected that in future devices Quantum Error Correction (QEC) will enable fault tolerant quantum computation with errors continuously corrected as a computation is executed.
However, successfully implementing QEC requires multiple, high-fidelity qubits to encode a single logical qubit. Although initial implementations of error correction codes have been demonstrated~\cite{acharya2022suppressing,ryan2022implementing}, QEC at a scale resulting in quantum advantage requires substantial further improvement in quantum hardware. Consequently, techniques reducing the impact of errors without performing QEC are crucial to obtain a near-term quantum advantage.

Error mitigation methods are techniques for reducing errors in near-term (non-fault-tolerant) quantum hardware. They can be applied on devices with larger error rates and smaller qubit numbers than required by QEC~\cite{cai2022quantum}. Various error mitigation techniques have been proposed, including dynamical decoupling~\cite{viola1999dynamical}, measurement error mitigation~\cite{maciejewski2020mitigation,nation2021scalable}, and noise-aware circuit compilation~\cite{cincio2018learning,cincio2021machine,murali2019noise,khatri2019quantum}.

\begin{figure}[ht]
    \includegraphics[width=0.99\columnwidth]{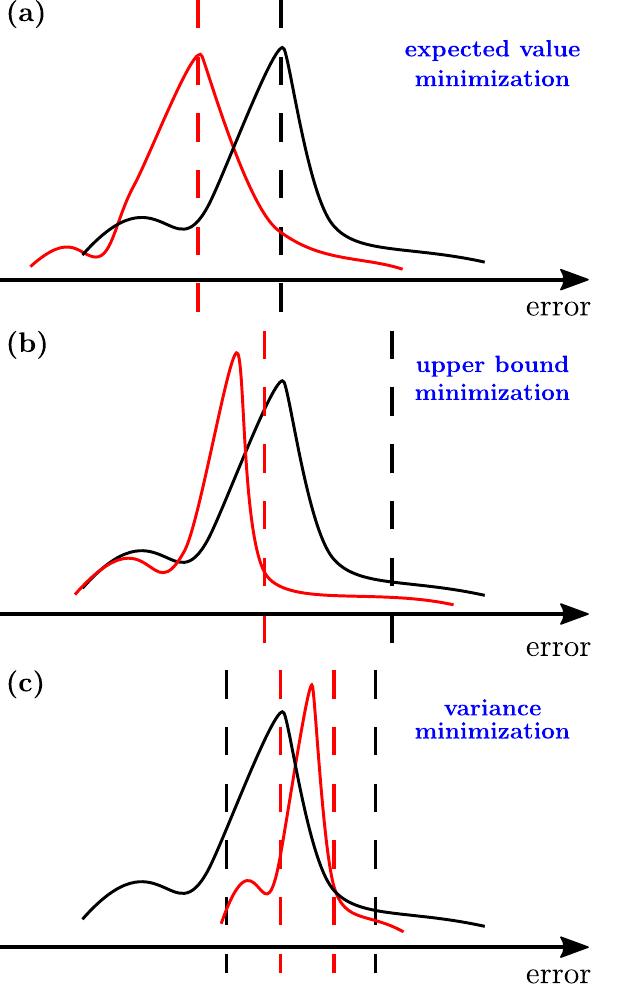}
    \caption{ {\bf Robust design of error mitigation under uncertainty (a cartoon depiction).} In this work, we present techniques aimed at enhancing the efficiency of error mitigation by considering the possibility of unknown probability distributions associated with error-mitigated observables. We propose methods to improve the distribution of error-mitigated outcomes with respect to key statistical quantities, through optimization of error mitigation hyperparameters (such as choice of ZNE noise levels or learning-based error mitigation training circuits).
Here we depict minimization of {\bf (a)} expected error, {\bf (b)} worst case error (tail value at risk), and {\bf (c)} variance of mitigated results, with the initial distribution in black, and the optimized one in red. The red and black dashed lines mark the initial and optimal values of the optimized quantities, respectively.
    }
        \label{fig:cartoon} 
\end{figure}

A widely-used approach to error mitigation aims to correct noisy expectation values of observables of interest with classical post-processing of measurement outcomes~\cite{cai2022quantum}. Examples of such methods are Zero Noise Extrapolation (ZNE)~\cite{temme2017error,kandala2018error,dumitrescu2018cloud,otten2019recovering,giurgica2020digital,he2020zero,cai2020multi,kim2021scalable}, Virtual Distillation~\cite{koczor2020exponential,huggins2020virtual,czarnik2021qubit,koczor2021dominant,huo2021dual, cai2021resourceefficient,seif2022shadow,hu2022logical,kwon2023efficacy} and learning-based error mitigation~\cite{czarnik2020error, strikis2020learning, montanaro2021error, vovrosh2021simple, urbanek2021mitigating, lowe2020unified, bultrini2021unifying}.
ZNE measures an observable of interest at multiple noise strengths and extrapolates it to the zero-noise limit. Virtual Distillation uses multiple copies of a noisy quantum state to ``distill'' its purer version suppressing incoherent errors. Classical post-processing of measurements of such purified states is used to obtain mitigated expectation values for observables of interest. Learning-based error mitigation uses classically simulable quantum circuits similar to a circuit of interest to train an ansatz that corrects effects of noise on expectation values of observables.
Among other approaches to error mitigation utilizing classical post-processing are quasi-probabilistic error decomposition~\cite{temme2017error,takagi2020optimal},  verified phase estimation \cite{o2020error}, truncated Neumann series~\cite{wang2021mitigating} and application-specific approaches leveraging symmetries of a circuit of interest~\cite{mcardle2019error,bonet2018low,otten2019noise,cai2021quantum}.

A fundamental limitation of the power of such error mitigation techniques is shot noise. Noisy expectation values are estimated using a finite number of state measurements called shots. Due to finite shot numbers the accuracy of these estimates is limited. This effect is called shot noise uncertainty. Shot noise uncertainty propagates through a classical post-processing procedure affecting error-mitigated expectation values.  It is well-known that error-mitigated observables typically have larger shot noise uncertainty than their noisy counterparts~\cite{endo2018practical}. Furthermore, for a wide class of error-mitigation protocols, the number of shots required for a given uncertainty grows exponentially with circuit depth fundamentally limiting the power of error mitigation~\cite{wang2021can,takagi2021fundamental,tsubouchi2022universal,takagi2022universal}. In the worst case, this growth is even faster~\cite{quek2022exponentially}.
Moreover, while error mitigation reduces bias caused by noise, it can introduce subtler biases. For example, ZNE performed with imperfect noise strength control or improper choice of extrapolation method can result in biased outcomes. Similarly, coherent errors in the case of Virtual Distillation and poor choice of training circuits for learning-based error mitigation produce bias.

Taking into account these limitations, it is crucial to account for the outcome uncertainty while applying and designing error mitigation methods to correct noisy observables. While methods to estimate shot uncertainty are known for some particular techniques~\cite{dumitrescu2018cloud,endo2018practical,cirstoiu2022volumetric}, no general approach to quantify the uncertainty of error-mitigated results or to optimize the robustness of the error mitigation under uncertainty have been proposed.
In this work, we fill this gap by introducing such methods, as shown schematically in Fig.~\ref{fig:cartoon}.
Our approach extends a rigorous uncertainty-quantification framework, which for any given set of assumptions and prior data derives optimal uncertainty bounds by solving optimization problems. The solutions of these problems give the extreme probabilities of failure or deviation under constraints of the relevant assumptions and the available data~\cite{owhadi2013optimal, sullivan2013optimal, kamga2014optimal}.
In particular, this framework is structured to not implicitly impose inappropriate assumptions, nor repudiate relevant information, and thus is well-suited for rigorous calculations of statistical quantities and optimal bounds on statistical quantities integral for the robust design of complex systems~\cite{mckerns2011building, mckerns2019optimal}.

We first introduce methods for robust design under uncertainty in Section~\ref{sec:UQ_intro}, and then provide some proof-of-principle demonstrations. We briefly introduce ZNE used in the manuscript in Section~\ref{sec:ZNE}, and then describe our error mitigation test-case setup in Sections~\ref{sec:setup_ZNE}. In Section~\ref{sec:UQ_results_ZNE} we demonstrate quantification of uncertainty for a ZNE-mitigated observable, while in Sections~\ref{sec:robust_design_ZNE}, \ref{sec:transfer} we demonstrate design of robust error mitigation for our benchmark cases.  We state conclusions in Section~\ref{sec:conclusions}. Additional examples of the applications of our methods, including
learning-based error mitigation, and technical details are provided in App.~\ref{app:surrogate_opt},~\ref{app:ZNE_diffev},~\ref{app:bootstrap},~\ref{app:CDR_num}.

\section{Robust design under uncertainty for error mitigation}
\label{sec:UQ_intro}

In this work, we consider the task of mitigating $O^{\rm noisy}$, a noisy, finite-shot estimate of an observable expectation value. Here, $O$ is an observable for a circuit of interest.
We assume that an error-mitigated estimate of this expectation value, $O^{\rm mitigated}$, is obtained by classical post-processing of the expectation values of noisy observables obtained from the circuit of interest or its modifications. As these noisy expectation values are estimated with a finite shot number, they are random variables that can be characterized by probability distributions. Consequently, $O^{\rm mitigated}$ is also a random variable characterized by a probability distribution,
$P(O^{\rm mitigated})$. To reliably quantify the uncertainty of $O^{\rm mitigated}$, we need to estimate properties of $P(O^{\rm mitigated})$.

To date, the only algorithms that have been proposed are those for estimating the variance and bias of $O^{\rm mitigated}$ (for some error-mitigation methods)~\cite{cirstoiu2022volumetric, govia2025bounding, mohammadipour2025direct}. Furthermore, these algorithms frequently assume that $P(O^{\rm mitigated})$ is a Gaussian distribution. This assumption is not valid for many error mitigation protocols. For instance, a ratio of Gaussian distributions gives rise to a Cauchy probability distribution, which is heavy-tailed, and does not have a well-defined variance~\cite{diaz2013existence}. This ratio of observables gives $O^{\rm mitigated}$ in the case of Virtual Distillation~\cite{koczor2020exponential}.

The most reliable way to calculate the expected values of statistical quantities of unknown distributions like $P(O^{\rm mitigated})$ is through sampling.
We propose using this approach to estimate the expected value and variance of $O^{\rm mitigated}$, and also to accurately capture the behavior at the tails of its distribution. Similarly, when the exact expectation value is known, the sampling of $P(O^{\rm mitigated})$ enables estimation of an error distribution.   
In this work, we demonstrate the feasibility of this approach to uncertainty quantification for error mitigation, by sampling a distribution of the relative error of $O^{\rm mitigated}$ defined as
\begin{equation}
\eta = 2\frac{|O^{\rm exact}-O^{\rm mitigated}|}{|O^{\rm exact}+O^{\rm mitigated}|},
\label{eq:model_score}
\end{equation} 
where $O^{\rm exact}$ is an exact expectation value of the observable for the circuit of interest.
Alternatively, the proposed methods can be straightforwardly applied to quantify variance and tail value at risk of $O^{\rm mitigated}$. In such cases, knowledge of $O^{\rm exact}$ is not necessary. Furthermore, other measures of error, like absolute error, can be used.   
A statistical characterization of $P(O^{\rm mitigated})$'s or $P(\eta)$'s distribution on average and in the tails, is crucial for understanding the uncertainty of the mitigation outcomes and enables us to design error mitigation methods that are robust under uncertainty.

An example of such a design is the optimization of error mitigation bias versus uncertainty for a given shot budget, $N_s^{\rm tot}$.  In particular, in the case of ZNE and a fixed $N_s^{\rm tot}$, higher-order polynomial extrapolation leads to a lower bias, while resulting in higher shot uncertainty~\cite{giurgica2020digital}. Similarly, in the case of Virtual Distillation, increasing the number of state copies results in better suppression of coherent errors and increased shot uncertainty~\cite{czarnik2021qubit}. In learning-based error mitigation, a similar effect arises with respect to the expressive power of the ansatz employed to correct noisy observables~\cite{bultrini2021unifying}.
The ZNE extrapolation method, Virtual Distillation copy number, or learning-based error mitigation ansatz choice are examples of error mitigation hyperparameters that can be adjusted to optimize performance or robustness.
While heuristic hyperparameter choices have been proposed~\cite{giurgica2020digital,czarnik2022improving}, no systematic methods for hyperparameter choice have been introduced so far.

In this work, we propose optimizing properties of $P(O^{\rm mitigated})$ -- or the distribution of its error -- with respect to error mitigation hyperparameters to maximize mitigation robustness under uncertainty.   We note that in order to quantify and minimize bias, a reference exact expected value is necessary, to minimize shot uncertainty across the hyperparameters no such knowledge is required. Furthermore, when an error mitigation method does not use special properties which make a quantum circuit simulable, we expect that hyperparameter values that minimize bias can be learned from classically simulable circuits and applied more broadly.   

As an example of robust design, we show below proof-of-principle minimization of a tail value at risk.
Intuitively, the tail value at risk is an expected value of the upper tail of $P(\eta)$  or $P(O^{\rm mitigated})$ distribution and quantifies the worst-case errors or the largest deviations from typical mitigation outcomes. More precisely, we define the tail value at risk for the error ${\rm TVaR}^{\rm right}_{\beta}$ using a $\beta$-quantile of the distribution $\eta^{\beta}$, which is the smallest value of $\eta$ for which
\begin{equation}
P( \eta \le \eta^{\beta}) \ge \beta,
\end{equation}
where  $P( \eta \le \eta^{\beta})$ is the probability of $\eta$ being smaller or equal to $\eta^{\beta}$. Then,
\begin{equation}
{\rm TVaR}^{\rm right}_{\beta} = E(\eta \,| \, \eta \ge \eta^{\beta} ) = \frac{1}{1 - \beta}\int_{\eta^{\beta}}^{\infty} \eta f(\eta) d\eta,
\label{eq:TVaR}
\end{equation}
where $E(\eta \,| \, \eta \ge \eta^{\beta} )$ is the expected value of $\eta$ under the condition $\eta \ge \eta^{\beta}$, and $f(\eta)$ is a probability density function of $P(\eta)$. For clarity, we often denote the tail value at risk by ${\rm TVaR}^{\rm right}_{\beta}(\eta)$. 
For given hyperparameter values, we estimate ${\rm TVaR}^{\rm right}_{\beta}$  by sampling error mitigation outcomes, as proposed above. In the spirit of a variational quantum algorithm, we run the ${\rm TVaR}^{\rm right}_{\beta}$ estimation as a cost function evaluation routine of a classical optimization algorithm \cite{cerezo2020variationalreview}. Using this approach, we obtain the hyperparameters that minimize ${\rm TVaR}^{\rm right}_{\beta}$, and hence maximize the worst-case error mitigation performance.   Furthermore, in App.~\ref{app:CDR_num} we demonstrate optimization of the error expected value $\langle \eta\rangle$, which characterizes the quality of the mitigated observable on average.

While we demonstrate our approach for Zero Noise Extrapolation and Clifford Data Regression in this manuscript, it has application to a broad class of error mitigation methods.  In particular, the proposed uncertainty  quantification framework can be applied to any error mitigation algorithm based on classical post-processing of quantum computation outcomes. Furthermore,  its shot cost  is determined by the distribution of the mitigated observable $P(O^{\rm mitigated})$, rather than by the size or depth of the circuit. 
Similarly, robust design can be applied whenever we have an error mitigation protocol  for which $P(O^{\rm mitigated})$ can be sampled and which has hyperparameters. In such cases, the optimization shot cost depends on  the properties of an optimization landscape, like the number of extrema or the magnitudes of the cost gradient.

\section{Proof-of-principle numerical results}

\label{sec:results}
In this section, we demonstrate the uncertainty quantification and robust design methods using the widely-used Zero Noise Extrapolation (ZNE) approach as a test case. We introduce ZNE in Sec.~\ref{sec:ZNE}. In Sec.~\ref{sec:setup_ZNE} we describe our choice of the mitigated circuit, observable, and the noise model.
Next, in Sec.~\ref{sec:UQ_results_ZNE} we apply our framework to rigorously quantify the worst-case errors of a ZNE-mitigated observable. Building on that, in Sec.~\ref{sec:robust_design_ZNE} we design our ZNE implementation to maximize its robustness by minimizing worst-case errors. In Sec.~\ref{sec:transfer} we show that this robust implementation can be applied to a wide class of similar circuits without the need for circuit-specific optimization. Additionally, in App.~\ref{app:CDR_num}, to demonstrate wide applicability of the proposed methods,  we show proof-of-principle results for Clifford Data Regression.

\subsection{Zero Noise Extrapolation}
\label{sec:ZNE}

Zero Noise Extrapolation (ZNE) is a popular error mitigation approach~\cite{temme2017error}, which has been successfully applied to circuits that pose a challenge to classical methods~\cite{kim2023evidence}. Its implementation requires an evaluation of the expected values of the observable of interest for multiple noise strengths, also called noise levels.   The mitigated expected value is obtained by extrapolating from these noisy expectation values to the zero-noise limit.

Here, we numerically simulate a ZNE implementation for a circuit of interest compiled to a native IBM gate set $\{{\rm CNOT}, \sqrt{X}, R_Z(\theta)\}$. We scale the noise strength with a widely-used gate identity insertion technique~\cite{dumitrescu2018cloud,he2020zero}. More specifically, for a noise level $k\in\{1,2,\dots,n\}$,  we replace each CNOT gate in the circuit of interest by $\lambda_k=2k-1$ CNOT gates. We assume that the noise level $k$ corresponds to scaling of the overall noise strength by a factor $\lambda_k$. This assumption is motivated by error rates of 2-qubit gates being much larger than error rates of single-qubit gates for current quantum computers.  

We treat $n$ as a hyperparameter of our ZNE implementation and consider $n\in\{4,5, \dots, 10\}$. For each $k$, we evaluate the noisy expected value of the observable of interest $y^{\rm ZNE}_k$. We fit $y^{\rm ZNE}(\lambda)$ with a third-degree polynomial which is subsequently used to extrapolate to the zero noise limit. Hence, $O^{\rm mitigated}$ equals the value of the fitted polynomial at zero. A larger $n$ provides more data points for the fitting procedure. At the same time, for a fixed shot budget $N_s^{\rm tot}$, larger $n$ results in a smaller number of shots per data point, and larger shot uncertainty of $y^{\rm ZNE}_k$. Consequently, the optimal value for the hyperparameter $n$ is not obvious, and thus an interesting test case for robust design methods.

Furthermore, when $O$ is a Pauli observable, the shot uncertainty of $y^{\rm ZNE}_k$ depends not only on the shot number but also on its expected value. Thus, a uniform distribution of shots among the noise levels does not result in uniform shot uncertainty for the data points. Therefore, the optimal choice of shot allocation under a fixed $N_s^{\rm tot}$ presents another interesting robust design problem.

Here, we test the impact of shot allocation using a parameterized shot distribution that assigns $N_s^{(k)}$ shots to the noise level $k$ according to:
\begin{equation}
    N_s^{(k)} =  \frac{2 N_s^{\rm tot}}{n} \left[ (1 - 2 \alpha) \frac{k}{n+1} + \alpha \right],
    \label{eq:shot_distr}
\end{equation}
with a parameter $\alpha \in [0,1]$. For this range of $\alpha$, we have $N_s^{(k)} >0$ and $\sum_k N_s^{(k)} = N_s^{\rm tot}$, where the choice of $\alpha=0.5$ produces shots distributed uniformly across the noise levels. For $\alpha < 0.5$, $N_s^{(k)}$ increases monotonically with increasing $k$, while for $\alpha > 0.5$, $N_s^{(k)}$ decreases monotonically with increasing $k$. For our ZNE implementation, we use $N_s^{\rm tot}=10^5$. The chosen functional form of $N_s^{(k)}$ balances simplicity of the shot allocation with the ability to  investigate shot allocation bias towards the low and high noise levels under a fixed shot budget. We remark, that though we choose this form for our proof-of-principle demonstration,  any shot allocation scheme with free parameters can be optimized within our robust design framework.

\subsection{The setup}
\label{sec:setup_ZNE}

As a test case, we use a circuit that prepares the ground state of a 6-qubit, one-dimensional XY model given by the Hamiltonian
\begin{equation}
H = \sum_{\langle i,j \rangle} X_i X_j +Z_i Z_j
\label{eq:H}
\end{equation}
with periodic boundary conditions, where $X$ and $Z$ are Pauli matrices and $\langle i,j \rangle$ denotes a pair of nearest-neighbor sites. For this circuit, we mitigate the expectation value of a two-site correlator $X_0 X_3$. To prepare the ground state, we use a hardware-efficient ansatz, with parameters found by classical optimization that match the ground state energy with an accuracy better than $10^{-13}$. The optimized circuit was then compiled into the native IBM gate set~\cite{kandala2017hardware}. The compiled circuit contained $60$ CNOTs.
Furthermore, to investigate dependence of the optimal $\alpha$ and $n$ on the choice of circuit, in Sec.~\ref{sec:transfer} we apply our framework to circuits obtained by modifying angles of the $R_Z(\alpha)$ gates in the original circuit. We describe the construction of the modified circuits in detail in Sec.~\ref{sec:transfer}.

We use the \textit{Qiskit} software package~\cite{qiskit}, to perform numerical
simulations, where we model the noisy CNOT and $\sqrt{X}$ gates using two-qubit and single-qubit depolarizing noise channels. The channels precede the noiseless gates in the Schr\"odinger picture. They act on two- and  single-qubit reduced density matrices, $\rho$, as
\begin{equation}
    \mathcal{N}_{\lambda}\left(\rho \right) = (1 - \lambda) \rho + \lambda \mathrm{Tr}\left[ \rho \right] \frac{I}{2^{n_q}},
\end{equation}
with $n_q=2$ and $n_q=1$, respectively, where $I$ denotes identity.  We choose the depolarizing error parameter $\lambda$ as $3.2 \cdot10^{-3}$ for the CNOTs, and $3.2 \cdot10^{-4}$ for the $\sqrt{X}$ gates. Furthermore, we assume that the $ R_Z(\alpha)$ gates are noiseless. These choices yield gate  error rates similar to the error rates of the real-world quantum computers.

\subsection{Uncertainty quantification}
\label{sec:UQ_results_ZNE}

We start with a demonstration of uncertainty quantification for our setup.
To quantify the uncertainty of the mitigated observable we perform error mitigation $N$ times. For each repetition,  we estimate the mitigated observable $O^{\rm mitigated}_i$ and its  error $\eta_i$ in Eq.~\eqref{eq:model_score}, where $i$ numbers the repetitions obtaining a sample of $N$ error-mitigated observable errors. Here, as a proof-of-principle application, we  use the sample to quantify behavior of the upper tails of the error distribution $P(\eta)$ by ${\rm TVaR}^{\rm right}_{\beta}$ in Eq.~\eqref{eq:TVaR}.
We estimate the tail value at risk from the sample as the mean of the sample elements larger than or equal to $\beta N$-th element of the sample sorted in ascending order.

\begin{figure}[t] 
    \includegraphics[width=0.99\linewidth]{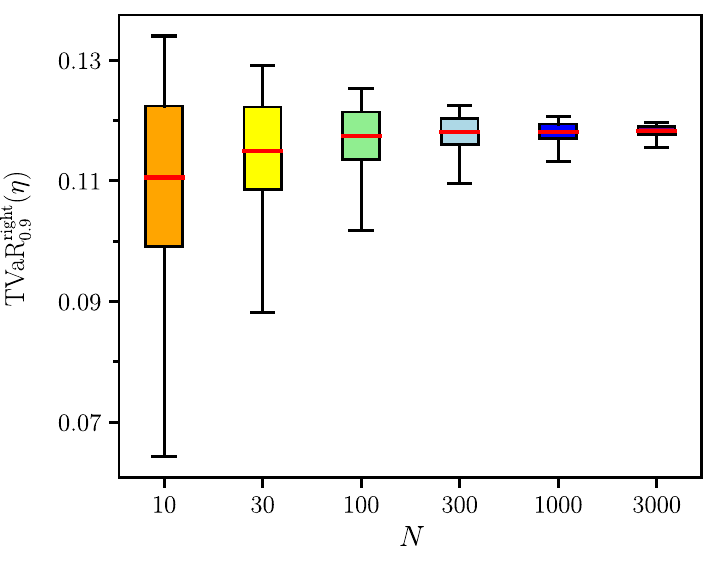}
    \caption{
    \textbf{Tail value at risk estimation for the ZNE error mitigation.} We estimate the tail value at risk ${\rm TVaR}^{\rm right}_{0.9}$ in Eq.~\eqref{eq:TVaR} of the mitigated observable relative error in Eq~\eqref{eq:model_score} by sampling error mitigation outcomes $N$ times. We plot distributions of the finite-$N$ tail value at risk estimates as boxplots. The red lines denote the medians of the distributions. The boxes correspond to ranges between the first ($Q_1$) and third distribution quartiles ($Q_3$). The whiskers denote the largest and the smallest data points. The estimates converge quickly with increasing $N$ indicating that  ${\rm TVaR}^{\rm right}_{0.9}$ can be used to reliably quantify the worst-case errors of the mitigated observable. Here, we estimate the distributions repeating ${\rm TVaR}^{\rm right}_{0.9}$ estimation $1000$ times for each $N$. We mitigate a ground state correlator $\langle X_0X_3 \rangle$ of a 6-qubit XY Hamiltonian in Eq.~\eqref{eq:H} under a depolarizing noise model.   ZNE is implemented with  $n=8$ noise levels, the total shot budget is $N_s^{\rm tot}=10^5$, and the shot allocation across the noise levels is determined by a distribution from Eq.~(\ref{eq:shot_distr}) with $\alpha=0.8$.}
    \label{fig:ZNE_TVaR} 
\end{figure}

In Fig.~\ref{fig:ZNE_TVaR} we plot the convergence of finite-$N$ estimates of ${\rm TVaR}^{\rm right}_{0.9}$ for an exemplary choice of $\alpha=0.8$ and $n=8$ with increasing $N$ for $N \in \{10,30,100,300,1000,3000\}$. As our quantity of interest is a random variable, its finite-$N$ estimate is also a random variable. To illustrate this, for each $N$  we generate 1000 samples of $O^{\rm mitigated}$ of size $N$ and estimate ${\rm TVaR}^{\rm right}_{0.9}$ from each sample. We find that such distributions of ${\rm TVaR}^{\rm right}_{0.9}$ estimates converge quickly with $N$.  This example demonstrates that reliable estimates of the tail value at risk can be obtained by sampling error mitigation outcomes. Moreover, in App.~\ref{app:CDR_UQ_results} we demonstrate $\braket{\eta}$ estimation, which quantifies the average error magnitude,  within our framework for the case of Clifford Data Regression.

\subsection{Robust design}
\label{sec:robust_design_ZNE}
Next, we demonstrate robust design by minimizing the ZNE worst-case error.
Specifically, we minimize the tail value at risk ${\rm TVaR}^{\rm right}_{0.9}$ of the ZNE relative error with respect to the hyperparameters $\alpha$ and $n$. We constrain $\alpha$ to $[0,1]$, and $n$ to $\left\{4, 5,\dots, 10 \right\}$.
To minimize the total shot cost of optimization, we restrict the number of ${\rm TVaR}^{\rm right}_{0.9}$ estimations to $30$ per optimization run, and we use an efficient surrogate-based minimization approach~\cite{o2025efficient}.
We initialize an optimization evaluating ${\rm TVaR}^{\rm right}_{0.9}$ for randomly chosen hyperparameters. We detail our surrogate optimization implementation in App.~\ref{app:surrogate_opt}.

\begin{figure}[t!] 
    \includegraphics[width=0.99\linewidth]{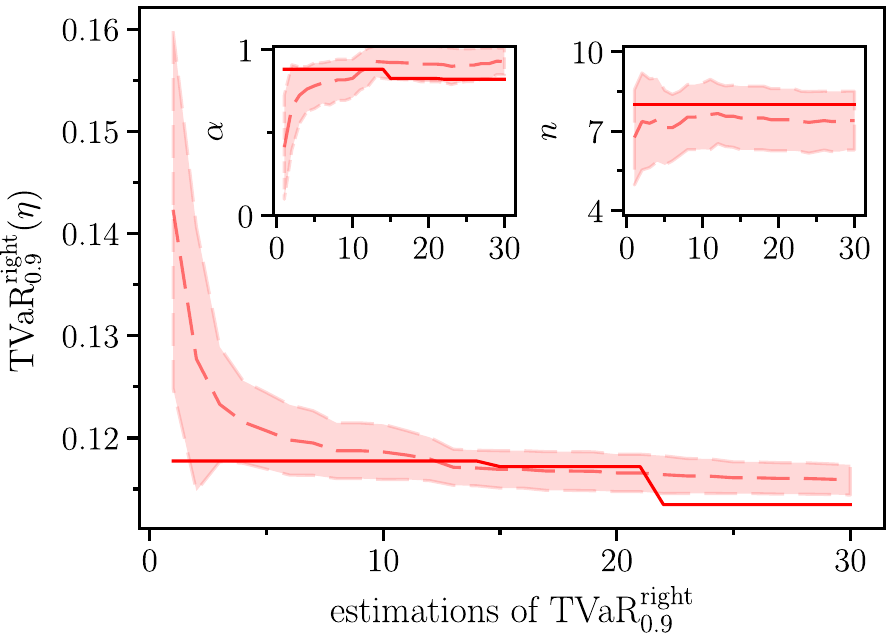}
    \caption{
    \textbf{Robust design of ZNE error mitigation.}
    We minimize the tail value at risk ${\rm TVaR}^{\rm right}_{0.9}$ of  the mitigated observable relative error  with respect to ZNE hyperparameters  $n$ (the number of the noise levels) and $\alpha$ (a parameter defining shot distribution across noise levels,  see Eq.~(\ref{eq:shot_distr})).  We perform $30$ optimization runs starting from random $n$ and $\alpha$ values. In the main plot, we show the convergence of the best ${\rm TVaR}^{\rm right}_{0.9}$ versus the ${\rm TVaR}^{\rm right}_{0.9}$ estimation number. The solid line shows the best optimization run. The dashed line shows convergence of ${\rm TVaR}^{\rm right}_{0.9}$ averaged over the runs, and the width of the shaded region is the standard deviation of ${\rm TVaR}^{\rm right}_{0.9}$ across the runs multiplied by~$2$. In the insets we use the same convention to  plot the convergence of the optimal $n$  and $\alpha$ estimates for the best run and averaged over the runs. The runs systematically improve  ${\rm TVaR}^{\rm right}_{0.9}$, and identify a hyperparameter space region with small ${\rm TVaR}^{\rm right}_{0.9}$. Furthermore, typically just several  ${\rm TVaR}^{\rm right}_{0.9}$ evaluations are enough to find this region.
    Here, as in Fig.~\ref{fig:ZNE_TVaR}, we mitigate $\left< X_0 X_3 \right>$ for the circuit preparing the ground state of the 6-qubit XY model subject to depolarizing noise. We estimate ${\rm TVaR}^{\rm right}_{0.9}$ from a sample of $N=1000$ error mitigation outcomes. ZNE is implemented with $N_s^{\rm tot}=10^5$.  The optimization runs are performed  under constraints $\alpha \in [0,1]$, and $n\in \{4,5,\dots,10\}$.
    }
    \label{fig:ZNE_optimization} 
\end{figure}

We perform $30$ optimization runs and investigate the best and average optimization run performance.  We gather the results in Fig.~\ref{fig:ZNE_optimization}.
We find that the optimization systematically improves ${\rm TVaR}^{\rm right}_{0.9}$. On average, the optimized ${\rm TVaR}^{\rm right}_{0.9}=0.116(2)$ substantially improves upon ${\rm TVaR}^{\rm right}_{0.9}=0.142(18)$ for a random hyperparameter choice within the optimization bounds. Here, the parentheses denote the standard deviations of the quantities.
Furthermore, an average run performs similarly to the best run, which yields ${\rm TVaR}^{\rm right}_{0.9}=0.113$. This indicates that the worst-case error can be reliably minimized with a small number of tail value at risk evaluations. Typically, just several ${\rm TVaR}^{\rm right}_{0.9}$ evaluations are enough to achieve the vast majority of the improvement.   The optimal hyperparameters averaged across the runs are $\alpha=0.93(8)$ and $n=7.4(1.1)$, and the best run gives $\alpha=0.82$, $n=8$.
Thus, they cluster in a region where more shots are assigned to lower noise levels, while the higher noise levels are included when performing the ZNE extrapolation.
In App.~\ref{app:ZNE_diffev}, we obtain very similar optimal parameter estimates with a more accurate and expensive differential evolution optimization. This provides further evidence that the tail value at risk can be reliably minimized with a small number of ${\rm TVaR}^{\rm right}$ evaluations. Here, we evaluate ${\rm TVaR}^{\rm right}_{0.9}$ using  $N=1000$  estimates of the error-mitigated observable. Consequently, the total shot cost of an evaluation-efficient optimization run is $3\cdot 10^9$. In App.~\ref{app:bootstrap}, we reduce it by a bootstrapping approach to $10^7$ while maintaining high quality of the optimized hyperparameters. We note that this shot budget is feasible for current superconducting quantum computers. Finally, we provide additional proof-of-principle examples of robust design for Clifford Data Regression in App.~\ref{app:CDR_robust_design}, where we target an optimization of the expected value of the error.  

\subsection{Generalization and transfer of the optimal hyperparameters.   }
\label{sec:transfer}
Reliable characterization of the mitigated observable distribution, as is required in robust design, typically uses many more shots than are required to perform error mitigation for the observable. 
Therefore, it is extremely useful to determine if the optimal hyperparameter values for a particular circuit can be transferred to other similar circuits. This question is motivated by an observation that for similar circuits the same error mitigation ansatz, that relates noisy and exact expectation values of observables, can be used for error mitigation~\cite{czarnik2020error, liao2023machine}. Hence, we expect that well-performing hyperparameters  should be transferrable to similar circuits.   Furthermore, for Pauli observables, the shot noise depends solely on expected values of the observables. Therefore, the shot uncertainty is similar for circuits with similar expected values.  To investigate \edit{the transferability issue}, we modify the ground state circuit by changing the angles of the rotation gates obtaining $20$ new circuits. We apply ZNE with $\alpha$ and $n$ learned for the original circuit in Fig.~\ref{fig:ZNE_optimization} to the new circuits and estimate  ${\rm TVaR}^{\rm right}_{0.9}$ of the mitigated observable. As above, we mitigate $X_0 X_3$. For a reference, we minimize the tail value at risk for each of the new circuits using the same optimization procedure as for the unmodified circuit.

\begin{figure}[t] 
    \includegraphics[width=1.0\linewidth]{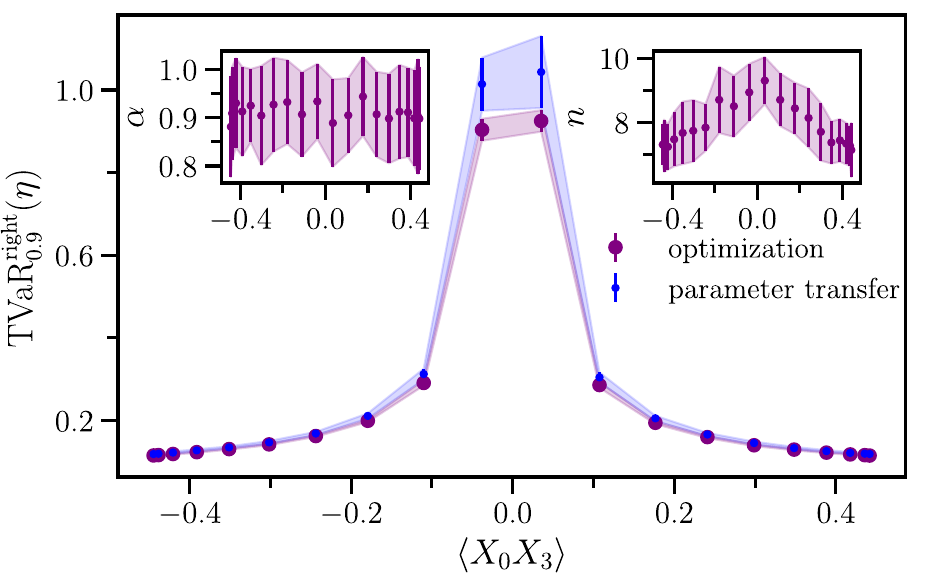}
    \caption{
    \textbf{Transfer of the optimal ZNE hyperparameters.}
    Here, we apply the optimal hyperparameters learned for the ground state circuit in Fig.~\ref{fig:ZNE_optimization} to  20 similar circuits. These circuits are obtained by modifying rotation angles of the original circuit. In all cases we correct $\left< X_0 X_3 \right>$.  In the main plot we show in blue ${\rm TVaR}^{\rm right}_{0.9}$ obtained for ZNE with the transferred hyperparameters   versus noiseless $\left< X_0 X_3 \right>$ of the circuits. For reference, in purple we show ${\rm TVaR}^{\rm right}_{0.9}$ for the circuit-optimized hyperparameters.  Furthermore, we plot these hyperparameters in the insets. Here, we transfer a set of $30$ hyperparameters obtained by the optimization runs from Fig.~\ref{fig:ZNE_optimization}. Similarly, for each circuit we run the optimization $30$ times to obtain optimal hyperparameter estimates.
    We plot ${\rm TVaR}^{\rm right}_{0.9}$ averaged across hyperparameters with dots and show the tail value at risk standard deviation with the error bars.
    We obtain similar  ${\rm TVaR}^{\rm right}_{0.9}$ values for the optimized and  transferred $n$ and $\alpha$. Furthermore, we observe that the distributions  of the optimized hyperparameters
    do not depend strongly on the circuit.
    }
    \label{fig:ZNE_parameter_transfer} 
\end{figure}

We find that both the transferred and the circuit-optimized hyperparameters result in very similar ${\rm TVaR}^{\rm right}_{0.9}$, as shown in Fig.~\ref{fig:ZNE_parameter_transfer}.  This clearly demonstrates the potential for hyperparameter transfer as an efficient mechanism for optimal circuit design.  In Fig.~\ref{fig:ZNE_parameter_transfer}, we plot the tail values at risk and the circuit-optimized hyperparameters versus circuits' $\langle X_0 X_3 \rangle$.  We find that the optimized $\alpha$ does not depend on the circuit choice, while the optimal $n$  grows slowly  with decreasing $|\langle X_0 X_3 \rangle|$. Nevertheless, only for the smallest $|\langle X_0 X_3 \rangle|=0.035$ the transferred angles perform visibly worse than the optimized ones. We note that in this case ZNE performs poorly as ${\rm TVaR}^{\rm right}_{0.9}$ is close to $1$ and the relative error of the observable of interest due to the shot noise is highest.  This may indicate that, while the hyperparameter transfer performs well on average, it may be further refined by accounting for special properties of the circuits of interest. We leave more detailed investigation of this issue to a follow-up work.

\section{Conclusions and discussion}
\label{sec:conclusions}
Error-mitigated observables exhibit uncertainty due to the propagation of shot noise variability from quantum measurements. This is one of the fundamental limitations of the power of error mitigation. Furthermore, error mitigation methods typically introduce bias that further limits the accuracy of error-mitigated results. In this work, we address these limitations by introducing methods to quantify and minimize the uncertainty and error of error-mitigated expectation values. Our error uncertainty quantification methods are generally applicable, in that they utilize unbiased sampling from a probability distribution of error-mitigated results (and the unbiased determination of bounds thereof). They enable one to estimate both expected and extremal values of error-mitigated observables, making it possible to quantify the robustness of error mitigation in a system. By applying this approach to classically simulable circuits, they can also be used to rigorously quantify the bias in error mitigation methods.

We leverage our uncertainty quantification methods to introduce robust design under uncertainty for error mitigation. By utilizing the optimization of hyperparameters, like the choice of noise levels in ZNE or CDR training circuits, one can fine-tune these hyperparameters to maximize the resilience of error mitigation to shot noise uncertainty and bias. In this work, we put forward a framework for the optimization of estimates of uncertainty and error of error-mitigated expectation values over error mitigation hyperparameters. This approach allows for systematic assessment of the sensitivity of error-mitigated results to the choice of hyperparameters and facilitates the identification of their best values, enhancing the potential of error mitigation.

Here we demonstrate both uncertainty quantification and robust design for test cases of ZNE and CDR error mitigation for correlators of the ground state of a 6-qubit one-dimensional XY model simulated with  depolarizing and IBM Toronto noise models, respectively. In the ZNE case, we estimate the  tail value at risk of the relative error of mitigated observables. Subsequently, we minimize this quantity with respect to hyperparameters controlling the choice of the noise levels and the distribution of shots across the noise levels.  Additionally, we demonstrate the potential for the efficient optimization through hyperparameter transfer from an original circuit to similar circuits.

We note that while this work showcases the feasibility of robust design for error mitigation methods, to fully realize its potential, further exploration of the methods performance and their enhancement is required. For example, an interesting application of our methods would be Virtual Distillation for which the estimates of the mitigated observable are heavy-tailed.   In particular, real-device noise is time-dependent.  Further study of the performance of our methods in such a case will be valuable for implementations on real-world quantum processors.  Another important goal for future research is to better understand the extent to which optimal hyperparameters can be generalized to modified circuits. In this work, we demonstrate the generalization of optimal hyperparameters to circuits that differ from the original circuit only by angles of the rotation gates. A natural extension is to determine whether the optimal hyperparameters can be generalized to a broader class of circuits.
Furthermore, in this work, we optimize the relative error of the mitigated observables. This choice requires knowledge of the exact expectation value of the mitigated observable. As the field moves towards error mitigation for circuits which are challenging for classical simulations~\cite{kim2023evidence}, a pertinent line of research is how to adapt these methods to such circuits. A promising direction is to reuse the optimal parameters obtained through robust design for near-Clifford circuits that differ from the target circuits only in their rotation angles.

\section{Acknowledgments}
We thank Fr\'ed\'eric Sauvage and Mike Martin for helpful conversations.
The research for this publication has been supported by a grant from the Priority Research Area DigiWorld under the Strategic Programme Excellence Initiative at Jagiellonian University.
PC acknowledges support  by  the National Science Centre (NCN), Poland under project 2022/47/D/ST2/03393. MM acknowledges support by the Uncertainty Quantification Foundation under the Statistical Learning program.
Research presented in this paper (ATS, MM) was also supported by the Laboratory Directed Research and Development (LDRD) program of Los Alamos National Laboratory under project number 20210116DR. The research was also supported (LC) by the Quantum Science Center, a National Quantum Science Initiative of the Department of Energy, managed by Oak Ridge National Laboratory.

\bibliography{bib_vieee}
\appendix

\section{Surrogate Optimization Algorithm}
\label{app:surrogate_opt}
For ZNE robust design, in Sec.~\ref{sec:robust_design_ZNE},~\ref{sec:transfer}, we employ surrogate-based optimization in a similar fashion to \cite{o2025efficient}, and adapt it to the case that one of the optimized hyperparameters was discrete.
At first, we randomly sample the initial $m_{\rm init}$ hyperparameter values $x_i = (\alpha_i,n_i)$ from a set $[0, 1]\times\{4,5,\dots,10\}$. Here, we use $m_{\rm init}=10$.  For each hyperparameter value $x_i$, we evaluate $y_i={\rm TVaR}^{\rm right}_{0.9}(x_i)$, performing error mitigation with $x_i$ hyperparameters.
We create a set of data points $\mathcal{D}=\big\{(x_i,y_i),\, i \in \{1,2,\dots,m_{\rm init}\}\big\}$.
Then we repeat $m_{\rm it}=20$ of the following steps:
\begin{enumerate}
    \item We build a classical surrogate of ${\rm TVaR}^{\rm right}_{0.9}$ as a function of $\alpha$ and $n$ using all data points in $\mathcal{D}$. The surrogate is constructed by  radial basis function interpolation of the data set $\mathcal{D}$. The interpolation is performed with the \textit{mystic} scientific machine learning package \cite{mckerns2009mystic,mckerns2011building}. The interpolation is performed under an assumption that $n$ is a continuous hyperparameter constrained to $n\in[4,10]$.

    \item We find a minimum of the classical surrogate obtained in the previous step using a \textit{mystic} implementation of a differential optimization algorithm. We denote the minimum by $\tilde x_{i}=(\alpha_i,\tilde n_i),\,i=m_{\rm init}+j$, where $j$ numbers the iterations.

    \item To obtain discrete $n_i$, we round $\tilde n_i$ to the closest integer. We take $x_i=(\alpha_i,n_i)$, and evaluate $y_i={\rm TVaR}^{\rm right}_{0.9}(x_i)$ from the error mitigation results. We add $(x_i,y_i)$ to the data set, $\mathcal{D} \to \mathcal{D}\cup\{(x_i,y_i)\}$.

\end{enumerate}
We estimate the optimal ${\rm TVaR}^{\rm right}_{0.9}$ as the minimum of the tail values at risk $y_i$  from $\mathcal{D}$, i.e., ${\rm min} \{y: \exists_x\, (x,y) \in \mathcal{D}\}$, and the optimal hyperparameters as the corresponding $x_i$.

\section{ZNE robust design with differential evolution algorithm} 
\label{app:ZNE_diffev}

Here, we compare performance of the surrogate-based ZNE robust design from Sec.~\ref{sec:robust_design_ZNE} to a more accurate and expensive robust design implementation utilizing a differential evolution optimization algorithm.  We choose the same setup as in Sec.~\ref{sec:ZNE}, namely we perform minimization of ${\rm TVaR}^{\rm right}_{0.9}$ for the ground state circuit and $X_0 X_3$ correlator with respect to hyperparameters  $\alpha\in[0,1]$ and $n\in\{4,5,\dots,10\}$.
We estimate ${\rm TVaR}^{\rm right}_{0.9}$ with $N=1000$ and have $N_s^{\rm tot}=10^5$.
We use \textit{mystic} implementation of differential evolution~\cite{storn1997differential} with optimizer hyperparameters set to \texttt{npop=20, maxiter=2000} minimizing likelihood of getting stuck in a local minimum. All optimization runs perform $40,000$ ${\rm TVaR}^{\rm right}_{0.9}$ evaluations. In comparison, the surrogate-based implementation from Sec.~\ref{sec:robust_design_ZNE} uses $30$ ${\rm TVaR}^{\rm right}_{0.9}$ evaluations per run.

\begin{figure}[t!] 
    \includegraphics[width=0.99\linewidth]{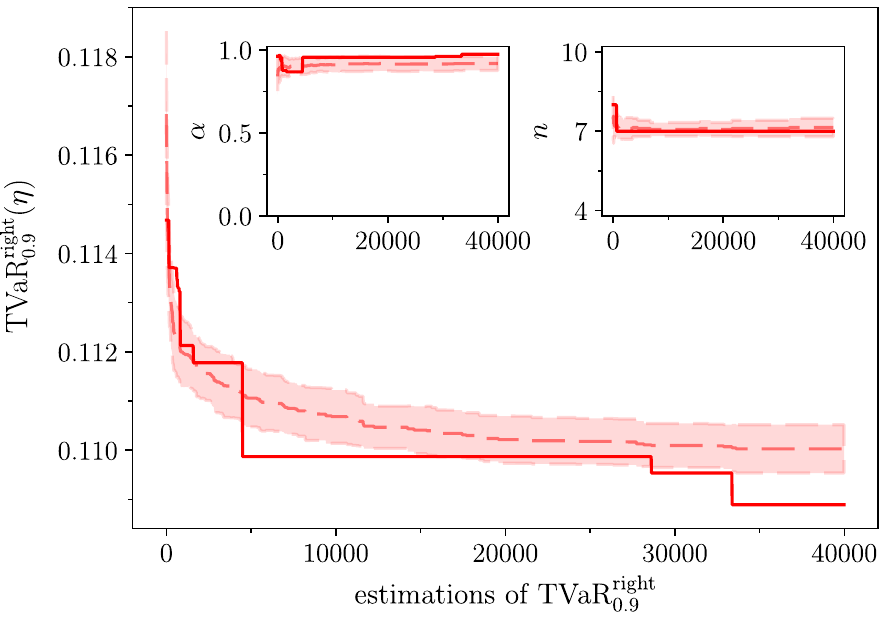}
    \caption{
    \textbf{Robust design of the ZNE error mitigation with  differential evolution optimization.}
    Here we  minimize the tail value at risk ${\rm TVaR}^{\rm right}_{0.9}$ of the relative error of the mitigated observable,  $\eta$, with respect to $n$  and $\alpha$ using  differential evolution optimization, which is more accurate and expensive than the surrogate-based optimization used in Fig.~\ref{fig:ZNE_optimization}. The mitigated circuit, observable, hyperparameter bounds, $N$, and $N_s^{\rm tot}$ are the same as in Fig.~\ref{fig:ZNE_optimization}.   In the main plot, we show the convergence of optimal  ${\rm TVaR}^{\rm right}_{0.9}$ versus the ${\rm TVaR}^{\rm right}_{0.9}$ estimation number for 28 optimization runs. The insets plot the convergence of the hyperparameters.  The solid lines show a run with the best final ${\rm TVaR}^{\rm right}_{0.9}$ value. The dashed lines show  $\eta$ and the hyperparameters averaged over the runs, and the shaded regions indicate  standard deviation of the error and the hyperparameters across runs. The optimal ${\rm TVaR}^{\rm right}_{0.9}$ and hyperparameter values are very similar to the ones obtained with the surrogate-based optimization from Fig.~\ref{fig:ZNE_optimization}, which uses three orders of magnitude less
      ${\rm TVaR}^{\rm right}_{0.9}$ estimations.  }
    \label{fig:ZNE_optimization_diffev} 
\end{figure}

We perform 28 optimization runs with random initialization. In Fig.~\ref{fig:ZNE_optimization_diffev} we show the best run and the results averaged across the runs. We find that the optimized ${\rm TVaR}^{\rm right}_{0.9}$ is very similar to the surrogate-based results from Fig.~\ref{fig:ZNE_optimization} both on average ($=0.1100(5)$ vs. $0.116(2)$) and for the best runs ($0.109$ vs. $0.113$), with differential evolution giving slightly lower results with smaller variance. Furthermore, the optimal hyperparameters averaged across runs agree within their standard deviations, as we obtain $\alpha=0.92(4)$, $n=7.1(4)$ with the differential evolution, and $\alpha=0.93(8)$, $n=7.4(1.1)$ with the surrogate-based method. Again, the differential evolution yields less variance of the estimates, as expected of a more accurate method.  Furthermore, the best runs yield similar optimal hyperparameters ($\alpha=0.97$, $n=7$, and $\alpha=0.82$, $n=8$, respectively). This shows that the efficient surrogate-based approach yields high-quality estimates of the optimal hyperparameters, and thus is a method of choice for real-world applications.

\section{Bootstrapping}
\label{app:bootstrap}

As the estimation of the statistical properties of the mitigated observable distribution requires multiple
repetitions of the error mitigation procedure, it typically requires large shot resources. Furthermore, robust design typically requires estimation of the distribution properties for multiple error mitigation hyperparameter values, which further increases the number of shots required. For example, the total shot cost of our ZNE optimization run (see Sec.~\ref{sec:robust_design_ZNE} for details) was $3\cdot 10^9$ shots. To improve the shot efficiency of the methods introduced here, a natural choice was to employ bootstrapping techniques. Such techniques generate additional data by resampling of the original data or a model based on the original data.

As a proof of principle, we numerically demonstrate robust design for our ZNE setup, see Sec.~\ref{sec:setup_ZNE}, while using a simple bootstrapping method. First, we spend in total $10^7$ shots to estimate the noisy expectation values $\langle X_0 X_3\rangle_{\rm est}^{(k)}$ for the ground state circuit and $k\in\{1,2,\dots,10\}$ noise levels. We divide the shots equally between the noise levels ($10^6$ shots per a noise level). Then, we use these estimates to model distributions of single-shot $\langle X_0 X_3\rangle$ measurements
\begin{equation*}
    P_{1}^{(k)} = \frac{1 + \left< X_0 X_3 \right>_{\rm est}^{(k)}}{2}  \quad {\rm and} \quad  P_{-1}^{(k)} = 1 - P_{1}^{(k)},
\end{equation*}
where $P_{1}^{(k)}, P_{-1}^{(k)}$ are probabilities of the $1$ and $-1$ outcomes for the noise level $k$, respectively. Here, we take into account that a distribution of single-shot measurements of a Pauli observable is a binomial distribution that is determined by the Pauli expectation value. Finally, we use the estimated distributions to generate results of the single-shot measurements for the robust design optimization runs.  Hence, this strategy required access to the quantum hardware only to estimate the distributions of the single shot noisy measurements, while the estimation of the distributions of the ZNE-mitigated observables was performed efficiently classically. Thus, the required quantum computing time is reduced by orders of magnitude.

\begin{figure}[t!]
    \centering
    \includegraphics[width=0.99\linewidth]{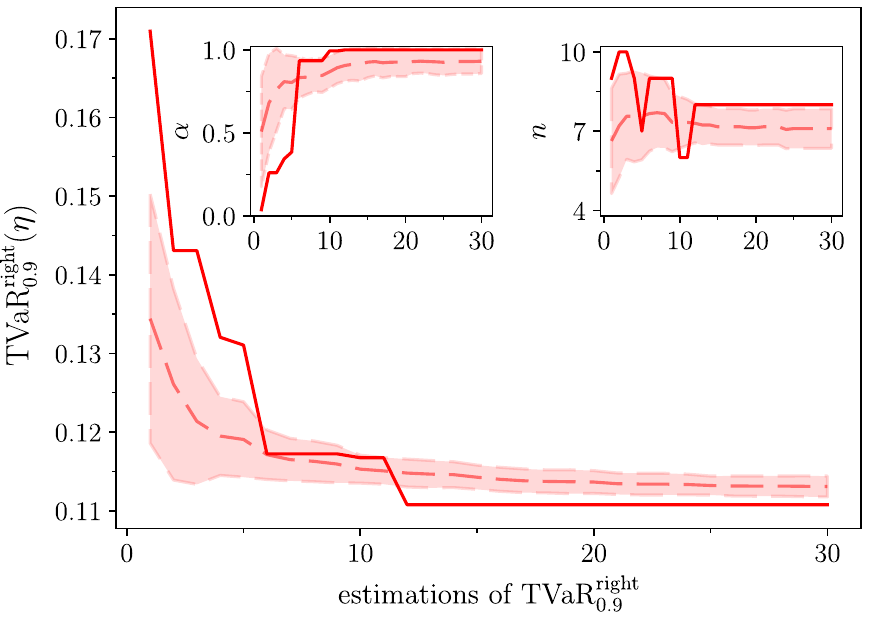}
    \caption{ {\bf ZNE robust design with bootstrapping.} 
    Here we  minimize the tail value at risk ${\rm TVaR}^{\rm right}_{0.9}$ of the relative error of the mitigated observable,  $\eta$, with respect to the  $n$  and $\alpha$ using  bootstrapping and the surrogate-based optimization. The usage of bootstrapping allows us to reduce  the total shot cost of an optimization run to $10^7$. For reference, the runs without bootstrapping that are shown in  Fig.~\ref{fig:ZNE_optimization} cost $3\cdot 10^9$ shots. The details of the bootstrapping optimization algorithm can  be found in App.~\ref{app:bootstrap}.   The mitigated circuit, observable, hyperparameter bounds, $N$ amd $N_s^{\rm tot}$ are the same as in Fig.~\ref{fig:ZNE_optimization}. Furthermore, we use the same surrogate-based optimization algorithm as in  Fig.~\ref{fig:ZNE_optimization}. In the main plot, we show the convergence of the optimal ${\rm TVaR}^{\rm right}_{0.9}$ versus the ${\rm TVaR}^{\rm right}_{0.9}$ estimation number for 30 optimization runs. The insets plot the convergence of the hyperparameters.  The solid lines show a run with the best final ${\rm TVaR}^{\rm right}_{0.9}$ value. The dashed lines show  $\eta$ and the hyperparameters averaged over the runs, and the shaded regions indicate  standard deviation of the error and the hyperparameters across runs.
    The optimal ${\rm TVaR}^{\rm right}_{0.9}$ and hyperparameter values are very similar to the ones obtained with the surrogate-based optimization from Fig.~\ref{fig:ZNE_optimization}. Hence, the reduction of the shot cost does not affect quality of the results.
    The best run has  $\alpha = 0.999, n = 8$, with ${\rm TVaR}^{\rm right}_{0.9} = 0.1108$. }
    \label{fig:ZNE_bootstrap}
\end{figure}

For our test case setup, the same as in the main text (see Sec.~\ref{sec:robust_design_ZNE} for details), we minimize ${\rm TVaR}^{\rm right}_{0.9}$ and constrain $\alpha$ and $n$ to $[0,1]$ and $\{4,5,\dots,10\}$, respectively. Furthermore, the same as in the case of the main text runs, we apply surrogate-based optimization as described in App.~\ref{app:surrogate_opt}, and choose the same $m_{\rm init}=10$, $m_{\rm it}=20$, $N=1000$, $N^{\rm tot}_s=10^5$.  We show the results of 30 optimization runs with bootstrapping in Fig.~\ref{fig:ZNE_bootstrap}, and find that they are very similar to the runs without bootstrapping from Fig.~\ref{fig:ZNE_optimization}, despite a reduction of the total number of shots required  by a factor of $300$. This example demonstrates the feasibility of huge shot-efficiency gains by the use of bootstrapping techniques.

\section{Proof-of-principle demonstration for Clifford Data Regression  }
\label{app:CDR_num}
 Here, we use Clifford Data Regression (CDR) to demonstrate usability of our uncertainty quantification and robust design methods beyond the ZNE approaches. Furthermore, to demonstrate a wide range of possible applications of our approach we consider here the expected value of the error $\braket{\eta}$ instead of the tail value at risk used in Sec.~\ref{sec:ZNE}, as our quantity of interest. We start with a short introduction of CDR in App.~\ref{app:CDR}. We follow that with a brief description of our test-case error mitigation problem in App.~\ref{app:CDR_setup}. We demonstrate the evaluation of $\eta$ within our framework in App.~\ref{app:CDR_UQ_results}, and robust design of CDR training circuits in App.~\ref{app:CDR_robust_design}.
 
\subsection{Clifford Data Regression}
\label{app:CDR}

Clifford Data Regression (CDR) is a learning-based error mitigation technique~\cite{czarnik2020error}. It uses classically simulable near-Clifford training circuits similar to the circuit of interest in order to correct a noisy expectation value of an observable of interest. Using CDR, we first find $N_t$  near-Clifford training circuits similar to a circuit of interest. Typically, one uses training circuits that differ from the circuit of interest only by gate rotation angles. Such training circuits can be obtained by substituting most  of the non-Clifford gates in the circuit of interest with Clifford gates of the same type~\cite{czarnik2020error}. For example, in the case of a rotation around the $z$-axis, $R_Z(\theta) = e^{-i (\theta/2) Z}$  with arbitrary $\theta$, one can replace this gate with a power of a phase gate, $S=e^{i\pi/4} e^{-i (\pi/4) Z}$, which is a Clifford gate. Here $Z$ is a Pauli operator. As long as the number of non-Clifford gates in a training circuit remains small enough, a Pauli expectation value can be efficiently simulated classically~\cite{fast2022pashayan}.
The Clifford substitutions  can be performed randomly~\cite{czarnik2020error}. Alternatively, they can be done with a Markov Chain Monte Carlo (MCMC) procedure to impose constraints on the expectation values of the training circuits~\cite{czarnik2022improving}.

Subsequently, we create training data by taking a pair of exact, $x_i \equiv O^{\rm exact}_i$, and noisy, $y_i \equiv O^{\rm noisy}_i$, expectation values of the observable of interest evaluated both classically and with a quantum computer, respectively. We do this for each training circuit. Here index $i$ enumerates training circuits. We fit the training data with a linear ansatz,
\begin{equation}
y_i = a x_i+b \ ,
\label{eq:CDR}
\end{equation}
where $a$ and $b$ are coefficients found by the least-squares linear regression. 
Finally, we use the resulting fitted coefficients to mitigate $O^{\rm noisy}$. We compute the mitigated expectation value as
\begin{equation}
O^{\rm mitigated} = a O^{\rm noisy} + b.
\end{equation}
Clifford Data Regression is based on the assumption that as long as the noise does not depend strongly on gate rotation angles, noise affects nearby circuits similarly, and therefore its effects can be learned from near-Clifford circuits. CDR has been found to match or outperform other state-of-the-art error mitigation methods while mitigating real-device noise~\cite{czarnik2020error, sopena2021simulating, cirstoiu2022volumetric}.
 
\subsection{The setup}
\label{app:CDR_setup}

We start with the same circuit of interest and observable of interest as was used in the main text with ZNE error mitigation. The circuit prepares the ground state of a 6-qubit XY model with periodic boundary conditions, and the mitigated observable is a two-site correlator $X_0 X_3$ --- see Sec.~\ref{sec:setup_ZNE} for a more detailed description of the circuit.
For this proof-of-principle demonstration, we performed noisy simulations using IBM's Toronto quantum computer noise model obtained using a built-in \textit{Qiskit}~\cite{qiskit} function that creates a noise model based on the device calibration.

In our tests, we use a modest shot number $N_s^{\rm tot}=10^4$ and $N_t=10$. We divide the shots between the noisy expectation values uniformly,
i.e.  for each $O^{\rm noisy}_i$, we use $N_s^{\rm tot}/(N_t+1)$ shots.
To mitigate the detrimental effects of shot noise, we previously proposed the use of training circuits with well-distributed exact expectation values of the mitigated observables, which can be generated by MCMC~\cite{czarnik2022improving}. Here, we apply uncertainty quantification to this technique. We use training circuits with $y_i$ uniformly distributed between $-y_{\rm max}$ and $y_{\rm max}$. These circuits are generated with the MCMC-based algorithm of Ref.~\cite{czarnik2022improving}. Our training circuits have $10$ non-Clifford gates, while the circuit of interest has $150$ non-Clifford gates. The MCMC procedure is initialized randomly, and produces samples of near-Clifford circuits with $y_i$ that are within $0.01$ from the desired $y_i$. Thus, the choice of training circuits is another source of uncertainty that potentially impacts $O^{\rm mitigated}$.

\begin{figure}[ht] 
    \includegraphics[width=\linewidth]{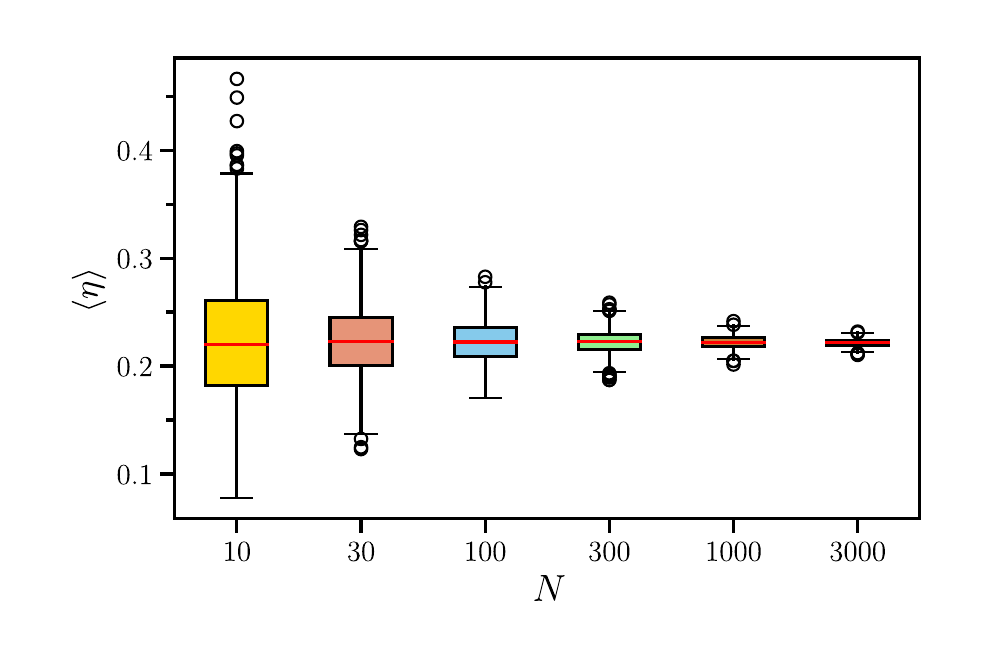}
    \caption{
    \textbf{The expected value of error quantification for CDR error mitigation.} Finite-sample estimates of  the mean  of the relative error in Eq.~\eqref{eq:model_score} of the CDR-mitigated  ground state correlator $\langle X_0X_3 \rangle$ of a 6-qubit XY model defined in Eq.~\eqref{eq:H}. For the purpose of examining the convergence of these estimates with the sample size, $N$, we plot their distribution for several values of $N\in\{10,30,100,300,1000,3000\}$. The distributions are plotted using boxplots. The red lines denote the medians of the distributions. The boxes correspond to ranges between the first ($Q_1$) and third distribution quartiles ($Q_3$). The whiskers denote the smallest and the largest datapoint within a range $[Q_1-1.5|Q_1-Q_3|,Q_3+1.5|Q_1-Q_3|]$. The circles denote outliers. For each value of $N$, $1000$ samples of size $N$ were generated to plot the distribution. Error mitigation is performed for an IBM Toronto noise model with hyperparameters $y_{\rm max}=0.5$ and $a=1$. }
    \label{fig:CDR_UQ} 
\end{figure}

To exemplify our robust design methods, we consider a more general form of the $y_i$ distribution parametrized by two parameters $y_{\rm max}$, and $a$. Namely, 
\begin{equation}
y_i = y_{\rm max} \, {\rm sgn}(r_i) |r_i|^a,
\label{eq:train_dist}
\end{equation} 
with $r_i$ values distributed uniformly in $[-1,1]$, and $\rm{sgn}$ denoting a sign function. $y_{\rm max}$  determines extreme values of the distribution, while $a$  determines a deviation from the uniform distribution. More precisely, $a=1$ corresponds to the uniform distribution, $a<1$ results in  clustering of $y_i$ around $|y_i|=y_{\rm max}$, and  $a>1$ causes clustering  around $0$.  This form of distribution systematically tests a heuristic strategy of distributing the training data proposed in Ref.~\cite{czarnik2022improving}.

\subsection{Uncertainty quantification}
\label{app:CDR_UQ_results}

We first demonstrate our uncertainty quantification approach applied to the expected value of the relative error $\eta$ defined in Eq.~(\ref{eq:model_score}). As in Sec.~\ref{sec:setup_ZNE},  we use a sample of $N$  error-mitigated observable estimates $O^{\rm mitigated}$ to robustly estimate $\braket{\eta}$ as an average of their errors, $\eta$.  In Fig.~\ref{fig:CDR_UQ}, we show convergence of a distribution of the finite-$N$ $\braket{\eta}$ estimates with an increasing $N$ plotted as boxplots. We consider $N=10-3000$, and for each $N$ generate $1000$ samples of size $N$. We choose $y_{\rm max}=0.5$, $a=1$. We find that the distributions converge rapidly as $N$ increases, making a reliable estimation of $\braket{\eta}$ possible.

We note that in sampling $O^{\rm mitigated}$, we sample both outcomes of quantum measurements and sets of training circuits consistent with hyperparameter values of $y_{\rm max}=0.5$ and $a=1$. Sampling the latter is done by sampling random starting points using the MCMC procedure. To minimize the classical cost of training circuit generation, we repeat the sampling of $O^{\rm mitigated}$ using precomputed  sets of $100$, $1000$, and $10000$  training circuits from which we randomly choose a set of training circuits used for CDR error mitigation. We observe that the distributions of finite sample $\braket{\eta}$ estimates are very similar for each choice, indicating that precomputed sets are sufficiently representative of the mitigated observable distribution and that the variance of error mitigation outcomes is due primarily to shot noise.

\begin{figure}[t] 
    \includegraphics[width=0.99\linewidth]{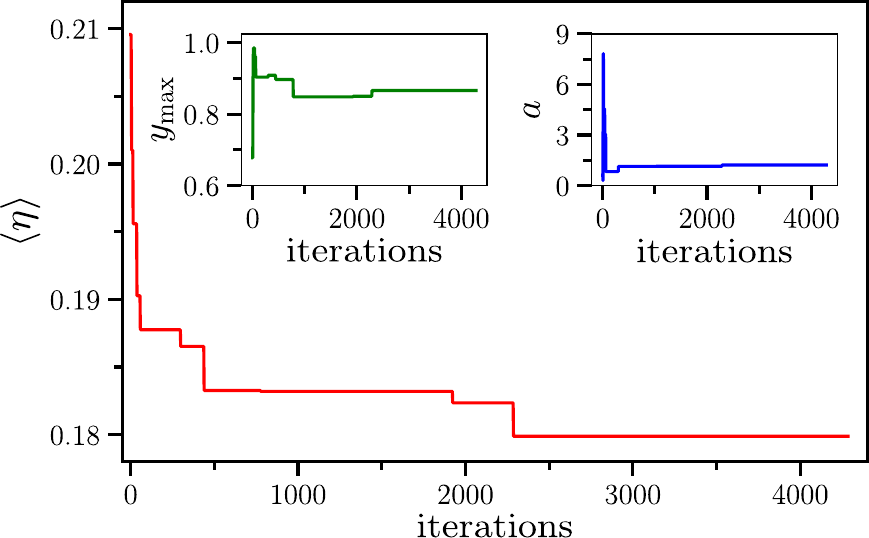}
    \caption{
    \textbf{Robust design of CDR training data to minimize the expected value of the relative error.} Optimization of exact expectation values of the CDR training circuits distribution in Eq.~\eqref{eq:train_dist}. Here we mitigate an $X_0X_3$ correlator of the ground state of a 6-qubit one-dimensional XY model defined in Eq.~\eqref{eq:H} for the IBM Toronto noise model. An estimate of the expected value of $\eta$ obtained with $N=1000$ samples is plotted versus a number of optimization iterations. The insets show the best hyperparameter values as a function of iterations.
     }   \label{fig:CDR_opt_EV}
\end{figure}

\subsection{Robust design}
\label{app:CDR_robust_design}
Here, we demonstrate robust error mitigation design by optimizing the hyperparameters $y_{\rm max}$ and $a$ to minimize $\langle \eta \rangle$. We perform a constrained optimization with $0.2 \le y_{\rm max} \le 1$ and $0.1  \le a  \le 10$, excluding values leading to extreme concentration of the training data.  As shown in Fig.~\ref{fig:CDR_opt_EV}, we obtain the best hyperparameter values $y_{\rm max}=0.87$ and $a=1.2$, corresponding to $\langle \eta \rangle =0.18$. These hyperparameter values result in well-distributed training circuit expectation values $y_i$, in agreement with a heuristic strategy proposed in Ref.~\cite{czarnik2022improving}.

\begin{figure}[H]
    \includegraphics[width=0.99\linewidth]{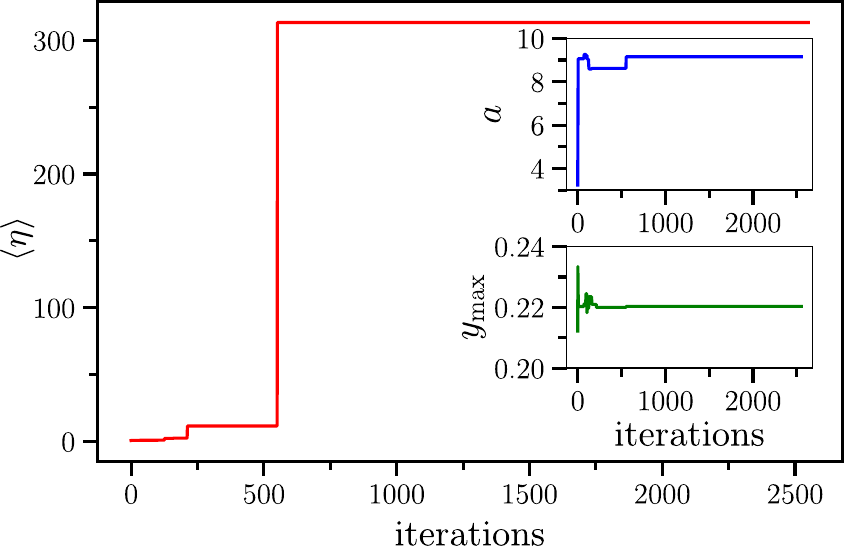}
    \caption{
    \textbf{Maximizing expected value of $\eta$.} 
     To investigate sensitivity of $\braket{\eta}$ to the CDR hyperparameters we  maximize $\braket{\eta}$ with respect to $y_{\rm max}$ and $a$. Here, the setup is the same as in Fig.~\ref{fig:CDR_opt_EV}.  In the insets (main plot), the  hyperparameter values maximizing the cost function (an estimate of  $\langle \eta \rangle$ obtained with $N=1000$) are plotted versus the optimization iterations.
     }   \label{fig:CDR_opt2_EV}
\end{figure}

Further, the framework can be used to rigorously investigate the sensitivity of error-mitigated results to the hyperparameters in Eq.~\eqref{eq:train_dist}. To demonstrate this application, we maximize the expected relative error with respect to $y_{\rm max}$ and $a$. We perform a constrained optimization with the same constraints on $y_{\rm max}$ and $a$, choice of the initial hyperparameter values, and value of $N$ as in App.~\ref{app:CDR_robust_design}. We obtained $\eta=310$ for $y_{\rm max} = 0.22$ and $a = 9.1$ as shown in Fig.~\ref{fig:CDR_opt2_EV}. This result  in combination with the results of $\braket{\eta}$ minimization demonstrates a strong dependence of the CDR performance on the training data hyperparameters $y_{\rm max}$ and $a$.  The parameters maximizing $\braket{\eta}$ correspond to the strongest clustering of training data around $0$ allowed by the constraints, confirming that the clustering negatively impacts the quality of error mitigation, as found in Ref.~\cite{czarnik2022improving}.

In this appendix, for the sake of clarity, we perform the optimizations using \textit{mystic}'s~\cite{mckerns2009mystic,mckerns2011building}  implementation of a differential evolution optimization algorithm~\cite{storn1997differential}, which is tuned to provide best result quality. Therefore, the optimization implementation used here is not shot-efficient. To minimize the detrimental effects of local minima, we perform the optimization starting from $9$ randomly chosen initial hyperparameter values and choose the optimization instance with the best expected value of $\eta$. For each pair of $y_{\rm max}, a$, the expected value of $\eta$ was estimated as the mean of a sample of $N=1000$ error mitigation outcomes.

\end{document}